\documentstyle[12pt]{article}
\font\titolo=cmbx10 scaled\magstep2

\font\tsnorm=cmr12

\font\tscorsp=cmti10

\def\NPB{{\em Nucl. Phys. }}
\def\PLB{{\em Phys. Lett. }}  

\def\PRD{{\em Phys. Rev. }}

\def\PRP{{\em Phys. Rep. }}
\def\IJMP{{ \em Int. J. Mod. Phys. }}
\def\z{Z\kern -4.6pt Z}
\def\dx{\int d^4x\ \sqrt{-g}\ }
\def\dxd{\int d^2x\ \sqrt{-g}\ }
\def\dxc{\int d^5x\ \sqrt{-g}\ }
\def\to{\rightarrow}
\def\lg{\left\{}
\def\lq{\left[}
\def\lt{\left(}
\def\rg{\right\}}
\def\rq{\right]}
\def\rt{\right)}
\def\ord#1{o\left(#1\right)}

\def\na{\nabla}
\def\a{\alpha}
\def\b{\beta}
\def\pa{\partial}

\def\e{\eta}
\def\f{\phi}
\def\F{\Phi}
\def\ef{e^{-2 \f}}
\def\eF{e^{-2 \F}}
\def\m{\mu}

\def\p{\psi}

\def\eqp{e^{-4\p}}
\def\r{\rho}
\def\s{\sigma}

\def\pu{\psi_{1}}
\def\pt{\psi_{3}}
\def\xd{\chi_{2}}
\def\xq{\chi_{4}}
\def\t{\tau}
\def\ds{ds^2=}

\def\l{\lambda}
\def\la{\l^2}

\def\be{\begin{equation}}
\def\ee{\end{equation}}
\def\bea{\begin{eqnarray}}
\def\eea{\end{eqnarray}}
\def\bc{\begin{displaymath}}
\def\ec{\end{displaymath}}
\def\lb{\label}
\def\q{\quad}

\begin{document}
\pagestyle{empty}
\null
\vskip 5truemm
\begin{flushright}
INFNCA-TH9904\\
\end{flushright}
\vskip 15truemm
\begin{center}
{\titolo  DIMENSIONAL REDUCTION OF 4D HETEROTIC }
\end{center}
\begin{center}
{\titolo STRING BLACK HOLES }
\end{center}
\vskip 15truemm
\begin{center}
{\tsnorm Mariano Cadoni}
\end{center}
\begin{center}
{\tscorsp Dipartimento di Fisica,  
Universit\`a  di Cagliari,}
\end{center}
\begin{center}
{\tscorsp Cittadella Universitaria, 09042, Monserrato, Italy,}
\end{center}
\begin{center}
{\tscorsp and}
\end{center}
\begin{center}
{\tscorsp  INFN, Sezione di Cagliari.}
\end{center}
\vskip 15truemm
\begin{abstract}
\noindent
We perform the spherical symmetric  dimensional reduction $4d\to2d$
of    heterotic string theory. We find a class 
of two-dimensional (2d) dilaton gravity 
models that gives a general description of the near-horizon, 
near-extremal behavior 
of  four-dimensional (4d) heterotic string black holes. We show that the
duality group of the 4d theory is realized in two dimensions  in 
terms of Weyl 
transformations of the metric. We use the 2d dilaton gravity theory 
to compute the statistical entropy of the near-extremal 4d, 
$a=1/\sqrt3$, black
hole.        
\begin{flushleft}
{PACS:  04.70.Bw; 11.25.Mj; 04.50.+h \hfill}\\
\end{flushleft}
\end{abstract}
\vfill
\hrule
\begin{flushleft}
{E-Mail: CADONI@CA.INFN.IT\hfill}
\end{flushleft}
\vfill
\eject
\pagenumbering{arabic}
\pagestyle{plain}
\section{Introduction}
\paragraph{}
The black hole solutions of four-dimensional (4d)  heterotic string theory
are very interesting from several points of view \cite{RD}.  They 
give  a general description of the whole set of charged black hole 
solutions of string theory (and general relativity) in four dimensions.
Heterotic, type $II A$ and type $II B$ strings in four dimensions  are 
connected 
one with the other by a web of dualities \cite{DLR,CA}, so that one 
can use the black hole solutions of 4d heterotic string theory as 
representative for the whole set  of 4d black hole solutions in string theory.
Moreover, using truncated models, it has been realized that there are only 
four  classes of black holes, which can be described by  effective  4d dilaton 
gravity models with dilaton  coupling $a=\sqrt 3, 1, 1/\sqrt 3, 0$ 
\cite{DLP,RA}.

The universal character of these solutions has found support both  from 
the compositeness idea, according to which the $a= 1, 1/\sqrt 3, 0$ 
solutions can be considered as bound states of the $a=\sqrt 3$ 
elementary solution \cite {RA} and from the interpretation of them as 
intersection of D-branes \cite{PT}. 

Each  class of 4d heterotic string  black holes 
is characterized by its spacetime structure, its singularities 
 and its thermodynamical 
behavior. 
Particularly remarkable is the  behavior of the 
$a=0$ black hole (essentially the Reissner-Nordstrom black hole of 
general relativity) compared with that of the $a= 1, \sqrt 3, 1/\sqrt 3,$  
cases.
The extremal limit of the $a=0$ black hole is   a  
zero-temperature  ground state with non vanishing entropy, whereas for
$a= 1, \sqrt 3, 1/\sqrt 3$ at the extremal limit we have zero temperature
and entropy. 
The previous features can be understood using  different approaches,
 even though  a complete and satisfactory picture of 
the all subject is still missing. For instance, the existence of
a ground state with the previously described behavior can 
be traced back to the presence of a mass gap. 
This mass gap can be explained both using string theory \cite{MS1} or just 
general features of the 4d effective dilaton gravity theory \cite{HW}.

Until now the various attempts to give a general description and 
to clarify the subject have mainly approached 
it from ``above'' , i.e using higher dimensional, $d\ge 4$, models. 
The recently 
proposed AdS$_{d}$ (anti-de Sitter)/ CFT$_{d-1}$ (conformal field theory)  
correspondence \cite{Wm} from one side and the discovery of dualities 
between four- and two-dimensional black holes \cite{Hy} from the other side,   
have changed  slightly the perspective.
For $d=2,3$ the AdS/CFT duality has been used to compute the 
statistical entropy of 2d  and 3d black holes \cite{St,CM}. On the other 
hand, it is known that near the horizon the geometry of the 4d, magnetically 
charged,  $a=0,1,1/{\sqrt  3}$ black hole factorizes as  the product of 
two spaces of constant curvature, ${\cal M}^{2}\times S^{2}$, where 
 $S^{2}$ is 
the two-sphere and ${\cal M}^{2}$ is 2d Minkowski space for $a=1$ and  
${\cal M}^{2}=AdS_{2}$ 
for $a=0,1/{\sqrt 3}$ \cite{CA}.
It is therefore natural to try to  extend  the  computation  of Ref. 
\cite {CM} of the 
statistical entropy to the 4d case. The hope is that the explanation 
of the entropy in terms of microstates will also  help to explain the  
different behavior of the various black hole solutions. 

Motivated by the previous arguments, in this paper  we perform  a 
generic $4d\to 2d$, 
spherical symmetric dimensional reduction of 4d  effective heterotic  
string  theory.   We find that the near-horizon, near-extremal behavior 
of 4d heterotic string black holes are described by a 
class of 2d dilaton gravity models. We show that these 2d models
encode all the relevant information about the 4d theory but in 
a much simpler form. 
In particular, we show that the 
duality group of the 4d theory is realized in the 2d theory in terms 
of  Weyl 
transformations of the metric and 
we use the 2d  dilaton gravity model
to compute the statistical entropy of the near-extremal 4d, 
$a=1/\sqrt3$,  black hole.

In Sect. 2 we describe the dimensional reduction $4d\to 2d$ of 
heterotic string theory and the single-scalar field truncation that 
produces the effective 2d dilaton gravity models. In Sect. 3 we  study 
the realization of the 4d duality symmetry in the 2d context,  showing 
that it corresponds to Weyl transformations of the metric. 
In Sect. 4 we use the 2d dilaton gravity model together with the 
results of Ref. \cite {CM} to compute the statistical entropy of the 
near-extremal, 4d, 
$a=1/\sqrt3$, black hole. Finally in Sect. 5 we draw our conclusions.

\section{ Dimensional reduction of  4d heterotic string theory}
\paragraph{}
In the string frame the bosonic action for heterotic string theory 
compactified on a six-torus \cite{MS,SE} can be written as follows:
\bea
 A_{H}&=&{1\over 16\pi}\dx \ef\lg R+4(\partial\f)^{2}- 
 2 \left[\left( \partial \s\right)^2+\left( \partial 
\r\right)^2\rq\right.\nonumber\\
&-& \left.{1\over 4}\lq e^{-2 \s- 2\r}F_{1}^{2}+e^{-2 \s+ 2\r}F_{2}^{2}+
e^{2 \s+2 \r}F_{3}^{2}
+ e^{2 \s-2 \r}F_{4}^{2}\rq\rg.
\lb{e1}
\eea
In the bosonic action of heterotic string theory 
toroidally compactified to $d=4$, we have set to zero the axion fields 
and all the $U(1)$ field strengths but four, two Kaluza-Klein fields
$F_{1}, F_{2}$ and  two winding modes  $F_{3}, F_{4}$. In the 
action (\ref{e1}) 
and throughout this paper we set the 
4d Newton constant $G=1$. The scalar fields $\f,\r, \s$ are related to 
the standard definitions of 
the string coupling, K\"ahler form and complex structure of the torus,
through  the equations
\be
\ef={\rm Im}S,\quad e^{-2\s}={\rm Im}T,\quad e^{-2\r}={\rm Im}U.
\lb{e2}
\ee
The extremal, Bogomol'ny-Prasad-Sommerfield (BPS), solutions of the action 
(\ref{e1}) are given by
(see for instance Ref. \cite{FK})
\bea\lb{e1a}
ds^{2}&=& - \pu \pt \, dt^{2}+\lt\xd\xq\rt^{-1} (dr^2+r^2d\Omega^2_{2}),
\nonumber\\
e^{4\f}&=&{\pu\pt\over \xd\xq},\quad e^{4\s}={\pu\xq\over \xd\pt},\quad   
e^{4\r}={\pu\xd\over \pt\xq},\nonumber\\
F_{1}&=&  \pm d\pu\wedge dt,\quad
\tilde F_{2}=\pm  d\xd \wedge dt, \quad
\quad F_{3}= \pm d\pt\wedge dt, \quad
\tilde F_{4}=  \pm d\xq \wedge dt, 
\eea
where $(\pu)^{-1}, (\pt)^{-1}, (\xd)^{-1},(\xq)^{-1}$ are  harmonic 
functions, $d\Omega^2_{2}$ is the  metric of the two-sphere and  
$\tilde{F}_{2}= 
 {e^{-2(\f+\s-\r)}}{^*}F_{2}$, 
 $\tilde{F}_{4}= {e^{-2(\f-\s+\r)}}{^*}F_{4}$
 ($^{*}$ denotes the Hodge dual).

Motivated by the fact that the action (\ref{e1}) admits  solutions 
that are 
the direct product  of two two-dimensional spaces of constant 
curvature ${\cal M}^{2}\times S^{2 }$ \cite{CA}, 
we study the  general, spherical symmetric, dimensional reduction  
$4d \to 2d$ of the 
theory.

Let us first fix our notation. Greek letters from the middle of 
the alphabet  denote 4d spacetime indices, $\mu ,\nu \ldots =0,1, 2,3$.
Greek letters from the beginning of 
the alphabet denote 2d spacetime indices, $\a,\b \ldots =0,1$.
The capital latin letters $I,J..= 2,3 $ label the coordinates of the 
internal two-sphere, $ S^{2 }$. The lower-case latin letters 
$i,j\ldots=1,2,3,4$ are used to label the four $U(1)$ field 
strengths $F_{(i)}$ whereas $a,b\ldots= 1,2,3$  label the 
moduli, $\e_{a}= (\f,\s,\r)$. 

The dimensional reduction of the 4d theory is performed by splitting 
the metric and the $U(1)$ field strengths into their 2d parts, using the 
ansatz
\bea
\lb{e3}
ds^{2}_{(4)}&=&ds^{2}_{(2)}+ Q^{2}e^{2\p} d\Omega^2_{2},\nonumber\\
F_{(i) \mu\nu}&=& \lg F_{(i) \a\b}, F_{(i) IJ}\rg,
\eea
where $Qe^{\p}$ is the 
radius of the two-sphere, $Q$ is a constant  that is related to the $U(1)$
charges (see Eq. (\ref{e6}) below)  and the scalar fields  $\e_{a},\p$ 
depend only 
on the 2d spacetime coordinates.
   
The 4d field equations together with the ansatz (\ref{e3}) constrain 
strongly the form of the fields  strengths $F_{(i)}$, we find
\be\lb{e4}
F_{(i)IJ}= p_{i}\epsilon_{IJ}, \quad  F_{(i) \a\b}= 
q_{i} e^{2 \e\cdot b_{i}-2\p}\epsilon_{\a\b}.
\ee 
In the previous equations, and in the following, the dot means
scalar product in the moduli space,   $ b_{i}$ are the vectors: 
\be\lb{e5}
b_{1}=(1,1,1), \quad b_{2}=(1,1,-1), \quad b_{3}=(1,-1,-1), \quad
b_{4}=(1,-1,1),
\ee
whereas $ p_{i}, q_{i}$ are respectively the
magnetic and  electric charge-vectors that characterize 
the particular dimensional reduction.
For sake 
of simplicity, we will consider only vectors of the form,
\be
\lb{e6}
 p_{i}=Q\hat{p_{i}}, \qquad  q_{i}=Q\hat{q_{i}},
\ee
where $\hat{p_{i}}$ and $\hat{q_{i}}$ are vectors with entries $0$ or $1$.

The form of $ \hat {p_{i}}$ and $\hat { q_{i}}$ determines the 4d 
background on which we are performing the dimensional reduction.
In general different charge-vectors will give rise to different  2d 
models. The 4d solutions are connected one with the other  by O(3,Z)  duality 
transformations, some of them change the spacetime structure of the 
solutions, others leave it invariant \cite {CA}. The most efficient 
way to organize the spectrum of the 4d solutions is to use the O(3,Z) duality 
symmetries together with the compositeness idea \cite{CA}.

It looks therefore very natural to  use the same procedure of Ref. 
\cite{CA} to classify 
the 2d models deriving from the dimensional reduction of the action 
(\ref {e1}). The 4d solutions can be classified in multiplets, labeled 
by $N$, of a given number (up to four) of elementary constituent, on 
which the duality symmetry $O(3, Z)$ acts in a natural way \cite{CA} . 
Moreover, the  multiplets  with $N=1,2,3,4$ can be put in correspondence 
with the 
solutions of the single-scalar, single $U(1)$ field strength model 
\cite {DLP,RA},
\be\lb{e6a}
A={1\over 16\pi}\dx \lg R -2(\partial\hat \f)^{2} - 
 {1\over 4} e^{-2 a\hat \f} F^{2}\rg,
\ee
with  coupling $a$ given respectively by $a=\sqrt 3, 1, 1/\sqrt 3, 0$. 

It turns out that 4d solutions characterized by the same number of 
magnetic and electric elementary constituents produce after 
dimensional reduction  the same 2d model. For this reason the various
dimensional reductions (or equivalently the corresponding 2d models) can be 
classified  by giving, apart form $N$, the numbers $n,m$  of electric, 
respectively,  magnetic elementary constituents, with the obvious constrain 
$N=n+m$. In this way one obtains 8 different 2d models, which can be 
put in correspondence with BPS states of the 4d model 
(\ref{e1}), 
\centerline{   }
\noindent
\leftline{$N=1$ multiplet, $a=\sqrt 3$,}
\bea\lb{e7}
n&=&1,\quad m=0,\quad \hat q_{i}=(1,0,0,0), \quad \hat p_{i}=0;\nonumber\\ 
n&=&0,\quad m=1,\quad \hat q_{i}=0,\q \hat p_{i}=(0,1,0,0).
\eea 
\noindent
\leftline{$N=2$ multiplet, $a=1$,}
\bea\lb{e8}
n&=&2,\quad m=0,\quad \hat q_{i}=(1,0,1,0), \quad \hat p_{i}=0;\nonumber\\ 
n&=&0,\quad m=2,\quad \hat q_{i}=0,\q \hat p_{i}=(0,1,0,1);\nonumber\\ 
n&=&1,\quad m=1,\quad \hat q_{i}=(1,0,0,0),\quad \hat p_{i}=(0,0,0,1).
\eea 
\noindent
\leftline{$N=3$ multiplet, $a={1\over \sqrt 3}$,}
\bea\lb{e9}
n&=&2,\quad m=1,\quad \hat q_{i}=(1,0,1,0), \quad \hat p_{i}=(0,1,0,0);
\nonumber\\ 
n&=&1,\quad m=2,\quad \hat q_{i}=(1,0,0,0),\q \hat p_{i}=(0,1,0,1).
\eea    
\noindent
\leftline{$N=4$ multiplet, $a=0$,}
\be\lb{f1}
n=2,\quad m=2,\quad \hat q_{i}=(1,0,1,0), \quad \hat p_{i}=(0,1,0,1).
\ee
In the previous equations we have given  only one representative 
element for each model, 
characterized by  $\hat p_{i}$ and $\hat q_{i}$.
There are, of course, different values of the U(1) charge-vectors that give 
rise to the same 2d model. For instance, the model with $N=1, n=1, 
m=0$  can be obtained not only with the values of $\hat q_{i}$ given 
in Eq. (\ref{e7}), but also  when $\hat p_{i}=0 $ and $\hat q_{i}$ has 
one entry equal to one, the others being zero.  
This degeneracy is related with the duality symmetries of the 
theory and will be discussed in detail in the next 
section.

In the following we will use   $S(N,m)$ to denote a 
dimensional reduction associated with a state with $N$ elementary 
constituents,
 of which $m$ are magnetic ($n=N-m$).
We can now perform explicitly the dimensional reduction of the model 
(\ref{e1}) defined by  the ansatz (\ref{e3}). The duality symmetry group
$O(3,Z)$ contains off-shell dualities, for this reason it is 
convenient to perform the dimensional reduction 
at the level of the equation of motion instead of 
directly dimensionally reducing the action (\ref{e1}).
Using the ansatz (\ref{e3}) and the equations (\ref{e4}) in the 4d field 
equations steming from the action (\ref{e1}), we get the following 2d 
field equations,

\bea\lb{f2}
R&+& 4\na^{2}\F -2 (\pa \F)^{2}-2 (\pa \p)^{2} -2 (\pa \e)\cdot(\pa 
\e)+\nonumber\\
&+&{1\over 2Q^{4}}e^{2\F}\lg \sum_{i}q_{i}^{2} e^{2 \e\cdot b_{i}}-
e^{-4\p}\sum_{i}p_{i}^{2} e^{-2 \e\cdot b_{i}}\rg+ {2\over 
Q^{2}}e^{-2\p}=0,
\eea
\be\lb{f3}
4\na_{\a}\lt\eF\na^{\a}\e_{a}\rt-{1\over Q^{4}}\lg \sum_{i}b_{ia}q_{i}^{2} 
e^{2 \e\cdot b_{i}}-
e^{-4\p}\sum_{i}b_{ia} p_{i}^{2} e^{-2 \e\cdot b_{i}}\rg=0, \quad 
a=2,3,
\ee
\be\lb{f4}
4\na_{\a}\lt\eF\na^{\a}\p\rt+{2\over Q^{4}}\eqp \sum_{i} 
p_{i}^{2} e^{-2 \e\cdot b_{i}}- {4\over Q^{4}}e^{-2(\p+\F)}=0,
\ee
   
\bea\lb{f5}
&&\na_{a}\na_\b\F +\pa_{a}\F\pa_\b\F - (\pa_{\a} \e)\cdot(\pa_{\a}\e)-
\pa_{a}\p\pa_\b\p+\nonumber\\
&& - g_{\a\b}\lg \na^{2}\F -{1\over 2} \lq(\pa \F)^{2}+
 (\pa \p)^{2} +(\pa \e)\cdot(\pa \e)\rq
 -{1\over 8Q^{4}}e^{2\F}\lt \sum_{i}q_{i}^{2} e^{2 \e\cdot 
 b_{i}}+\right.\right.\nonumber\\
&&\left.\left.e^{-4\p}\sum_{i}p_{i}^{2} e^{-2 \e\cdot b_{i}}\rt+ {1\over 
2 Q^{2}}e^{-2\p}\rg=0,
\eea
where the curvature $R$ and all the differential operators are
2d quantities,  the field $\F$ is the 2d dilaton related, as usual, 
to the 4d 
dilaton $\f$ by
\be\lb{f6}
\f=\F-\p,
\ee
and the vector $\e_{a}$ is now defined in terms of $\F$, 
$\e_{a}=(\F,\s,\r)$. 

The field equations (\ref{f2})-(\ref{f5}) are rather complicated. 
They assume a much simpler form by considering a consistent, 
single-scalar  
field truncation that reduces them to those of 2d gravity coupled to the 
dilaton $\F$.  These models have been widely investigated in recent years 
under the name of 2d dilaton gravity.
The single-scalar field truncation is obtained using an ansatz 
expressing the fields $\p,\s,\r$ in terms of the dilaton $\F$ in a way 
that is consistent with the field equations (\ref{f2})-(\ref{f5}),
\bea\lb{f7}
e^{2\p}&=&A^{2} \exp\lt {{4(m-2)\over a^{2}N}\,\F}\rt,\nonumber\\
e^{2\s}&=&A^{-1} \exp\lt {{2C_{\s}\over a^{2}N}\,\F}\rt ,\nonumber\\
e^{2\r}&=&A^{-1} \exp\lt {{2C_{\r}\over a^{2}N}\,\F}\rt,
\eea
where $A=\sqrt{3/2}$ for $N=1$, $A=1$ for  $N=2,3,4$.
$C_{\s}=-1$ for the states $S(1,1), S(2,1),$ $S(3,2)$;
$C_{\s}=0$ for $S(2,2), S(2,0), S(4,2)$;
$C_{\s}=1$ for $S(1,0), S(3,1)$.  
$C_{\r}=-1$  for the states $S(1,1), S(1,0), S(3,2),$ $ S(3,1)$;
$C_{\r}=0$ for $S(2,2), S(2,0), S(2,1),$  $ S(4,2)$.

After inserting Eqs. (\ref{e7})-(\ref{f1}) and (\ref{f7}) into the field 
equations (\ref{f2})-(\ref{f5}), we find that, for $N=1,2,3$,  they are 
equivalent to those 
derived from a class of 2d dilaton gravity models whose action has the 
form, 
\be \lb{f8}
 A_{2}={1\over 2}\dxd \eF\lq R+\Omega (\partial\F)^{2}+\la V(\F)\rq,
 \ee 
where  $\la=1/Q^{2}$ for $N=1,2$, and  $\la=1/(2Q^{2})$ for $N=3$.
The kinetic coefficient  $\Omega$ and the dilaton potential $V$ are 
completely determined in terms of $n, m,a, N,$
\be\lb{f8a}
\Omega = 8\lt {1-n\over a^{2 }N}\rt, \qquad V(\F)= \exp\lq 4\lt 
{2-m\over a^{2}N}\rt\F\rq.
\ee

For $N=4$ the field equations (\ref{f2})-(\ref{f5}) become equivalent to those 
derived from the action,
\be\lb{f9}
A={1\over 2}\dxd \eF\lq R+4(\partial\F)^{2}- {1\over 2} F^{2}+\la \rq,
\ee
which describes  the heterotic string in 2d target-space \cite{NY}.

The class of dilaton gravity models defined by the action (\ref{f8}) 
contains, as particular cases, models that have been already investigated 
in the past. For  $N=2, m=2$  the action describes the 
Callan-Giddings-Harvey-Strominger (CGHS) model \cite{CGHS}.  For $N=2, 
m=1$ we get 
the Weyl rescaled CGHS model investigate in Ref. \cite {CM3}. 
Finally, for $N=3, 
m=2$ we obtain  the Jackiw-Teitelboim (JT) model \cite{JT}.

In general the 2d dilaton gravity model (\ref{f8}) give a near-horizon 
description of the, near-extremal, 4d black hole solution of the action 
(\ref{e1}). 
This near-horizon description corresponds to 
setting to zero the constant term in the harmonic functions  
appearing in Eq. (\ref{e1a}). 
On the other hand, one can easily verify that  this is exactly the  way to 
make the ansatz 
(\ref {f7})  consistent   the 4d extremal solutions (\ref{e1a}).

\section{Dualities and Weyl transformations}
\paragraph{}

In the canonical frame $g_{C}= e^{-2\f}g_{S}$, the 4d field equations 
following from the action (\ref{e1}) are invariant under the action  
the 
$O(3,Z)$ duality group 
that acts on the moduli $\f,\s,\r$  and on the $U(1)$ field strengths but 
leaves the 4d metric unchanged \cite{CA}. The $O(3,Z)$ duality  group 
can  be generated  using  three $S-T-U$ duality transformations 
$\t_{S},\t_{T},\t_{U}$ together with the permutation group $P_{3}$ 
acting on the moduli  $\f,\s,\r$ and on the $U(1)$ field 
strengths
$F_{i}$ (see Ref.\cite{CA}). After 
passing to the string metric $g_{S}$ and performing the 
dimensional reduction described in the previous section, 
the duality group $O(3,Z)$ becomes a symmetry  of the 2d field 
equations (\ref{f2})-(\ref{f5}). Because the 4d action (\ref{e1}) 
is written in 
terms of the 
string metric the duality transformation will act also on the 2d metric 
$g_{\a\b}$ and, owing to Eq. (\ref{e4}), also on the
charge-vectors $p_{i}, q_{i}$. One can easily verify that the  O(3,Z) 
duality 
transformations leave invariant the 2d dilaton field $\F$, whereas 
in terms of the remaining  2d fields $\p,\r,\s,g_{\a\b}$ and of 
the charge vectors $p_{i}, q_{i}$ 
it is realized as follows:
\bea\lb{g1}
\t_{S}&:& \p\to -\p -2\F,\q  g_{\a\b}\to e^{-4 (\F+\p)}g_{\a\b},
\q q_{1}\to p_{3},\q 
q_{3}\to p_{1},\q q _{2}\to p_{4},\nonumber\\
& & q_{4}\to p_{2};\nonumber\\ 
\t_{T}&:& \s\to -\s,\q q_{1} \leftrightarrow q_{4},\q
 q_{2}\leftrightarrow q_{3};\nonumber\\
\t_{U}&:& \r\to -\r,\q q_{1}\leftrightarrow q_{2},\q 
q_{3}\leftrightarrow q_{4};
\eea
\bea\lb{g2}
P_{3}: & &\s \leftrightarrow  \r,\q  q_{2}\leftrightarrow q_{4};
\nonumber\\
& &\p\to  \s-\F,\q \s\to \p +\F,\q  g_{\a\b}\to e^{2 ( \s -\F-\p)}g_{\a\b},
\q  q_{3}\to p_{4},\q  q_{4}\to p_{3};
\nonumber\\
& &\p\to \s-\F,\q\s\to \r,\q \r\to \F+\p,\q g_{\a\b}\to 
e^{2 ( \s -\F-\p)}g_{\a\b},
\q q_{4}\to q_{2}, \nonumber\\
& &q_{2}\to p_{3},\q
q_{3}\to p_{4};
\nonumber\\
& &\p\to \r-\F,\q \r\to \F+\p,\q g_{\a\b}\to e^{2 ( \r -\F-\p)}g_{\a\b}, 
\q q_{2}\to p_{3}, \q
q_{3}\to p_{2};
\nonumber\\
& &\p\to \r-\F,\q\s\to\F+\p,\q \r\to \s,\q g_{\a\b}\to 
e^{2 ( \r -\F-\p)}g_{\a\b},
\q q_{2}\to q_{4}, \nonumber\\
& &q_{3}\to p_{2},\q
q_{4}\to p_{3}.
\eea

The previous equations describe the action of the duality group 
on  electric states ($p_{i}=0$), the action on magnetic states 
($q_{i}=0$) can be easily obtained from Eqs. (\ref{g1}),(\ref{g2}),
by interchanging $q_{i}\leftrightarrow p_{i}$.
The duality group generated by the transformations (\ref{g1}),(\ref{g2})
becomes extremely simple once we perform the single-scalar field 
truncation described in Sect. 2. One can easily verify that the 
transformations  $\t_{T},\t_{U}$ and the first transformation in Eq.
(\ref{g2}) change the ansatz (\ref{f7}) but not the resulting 2d 
action (\ref{f8}), (\ref{f9}). This fact has a natural explanation.
 $\t_{T},\t_{U}$
and the first duality in  Eq. (\ref{g2})
do not change the number $n$ of electric (or magnetic $m$) elementary 
constituents of a state, so that they cannot change the 2d action 
because the latter is parametrized in terms of $n$ and $m$ only.
On the other hand,  using Eqs. (\ref{f7}) into Eqs. 
(\ref{g1}),(\ref{g2}), one finds that the  
remaining duality transformations (in particular the $\t_{S}$ duality)  
act on the 2d dilaton gravity models
(\ref{f8}) and (\ref{f9})  as Weyl transformations of the 2d metric,
\be
\lb{g3}
g_{\a\b}\to e^{P\F} g_{\a\b}, \q P={4\over a^{2}N} \lt m-m'\rt,
\ee
which map one into the other  models in Eqs. (\ref{f8}), (\ref{f9}) 
with the same value of $N$ 
but with  a number $m$ and $m'$ of magnetic elementary constituents.

Acting at $N$ fixed, the Weyl transformations (\ref{g3}) connect one 
with the other  models within a given multiplet. 
For instance, taking $N=2$ we have three models:
$S(2,2)$ ($\Omega = 4$, $V(\f)=1$, the CGHS model), $S(2,1)$ ($\Omega = 0$, 
$V(\f)=\exp (2\F)$, the Weyl-rescaled  CGHS model of Ref. \cite{CM3} and 
$S(2,0)$ ($\Omega = -4$, 
$V(\f)=\exp (4\F)$). These models are obtained one from the other using 
the Weyl transformations (\ref{g3}). $g_{\a\b}\to e^{2\F} g_{\a\b}$, 
maps   $S(2,2)\to S(2,1)$ and $S(2,1)\to S(2,0)$ whereas 
$g_{\a\b}\to e^{4\F} g_{\a\b}$ maps  $S(2,2)\to S(2,0)$.

The $\t_{S}$ duality is essentially an electro/magnetic duality, so 
that the strong/weak coupling duality of the 4d theory 
(\ref{e1}) is translated, after dimensional reduction to two 
dimensions,  at least for the 4d model under 
consideration, into  a 
dilaton-dependent Weyl 
rescaling of the 2d metric.  

It has been shown that dilaton-dependent Weyl transformations leave 
invariant the physical parameters (the mass and for 2d black hole 
solutions, the Hawking temperature and radiation flux) associated with 
the solutions of 2d dilaton gravity \cite{CA2}. 
Also other physical parameters that can be expressed in terms of the 
mass and the temperature (e.g the entropy) are invariant under such 
transformations. 
The dimensional reduction 
seems to wash out most of the information about the magnetic or 
electric character of the 4d solution. At the 2d level the relevant 
information is encoded in the number $N$, the number of elementary 
constituents, the actual number of magnetic and electric elementary 
constituents being  irrelevant for the physical parameters of the 
solution.
At the level of the 4d theory this implies the equivalence of electrically and 
magnetically 
charged configurations, as long as only excitations near 
extremality are concerned. Moreover, in the 2d context, the duality 
implies the equivalence of spacetime structure that behave rather 
differently. 2d  spacetimes of constant curvature ( e.g  the solutions 
of the 2d model $S(3,2)$) are dual (i.e connected by Weyl 
transformations) to spacetimes with singularities (e.g the solutions
of the 2d model $S(3,1)$). 
  
Presently, we do not know if this is a peculiarity of the 4d 
heterotic string theory (\ref{e1}) or a general feature of 4d effective 
string theory. However, our results indicate that the dimensionally 
reduced 2d theory takes care only of the relevant physical properties of
the 4d model and can therefore  be used to give a universal 
classification  of the near-extremal behavior of the  4d black hole 
solutions of string theory.

\section{  Statistical entropy of the  a=1/${\bf\sqrt 3}$ 4d black hole} 
\paragraph{}
The results of the previous sections together with those of Ref. \cite{CM} 
can be used to calculate, 
microscopically, the entropy of the near-extremal 4d, $a=1/\sqrt3$, black hole. 
The near-extremal, near-horizon behavior of this 
black hole is described by the 2d model of Eq. (\ref{f8}) with 
$N=3, m=2$, i.e by the JT dilaton gravity model. 
For the JT   black hole  a derivation of 
the statistical entropy  has been given in Ref. \cite{CM}, one can, therefore,
use it to compute, microscopically, the entropy of the 
4d,  $a=1/\sqrt3$, black hole.

For $a=1/\sqrt3$ the single-scalar field model of Eq.(\ref {e6a})
takes  the form,
\be\lb{h1}
A={1\over 16\pi}\dx \lg R-2(\partial\hat \f)^{2}- 
{1\over 4}  e^{-{2\over \sqrt 3}\hat \f}F^{2}\rg,
\ee
where the scalar field $\hat \f$ is connected to the 4d dilaton $\f$ 
trough $ \hat \f= \sqrt 3 \,\f$. 
As mentioned above, this model arises as single-scalar field, single $U(1)$
field strength truncation of the $N=3$ composite solutions of the action 
(\ref{e1})  and as compactification of the five-dimensional (5d) 
Einstein-Maxwell theory.
The general (non extremal) black hole solution of the model has the 
form \cite{DLP1},
\bea\lb{h2}
ds^{2}&=&- H^{-{3\over2}}\lt 1-{\m\over r}\rt dt^{2}+ 
H^{3\over2}\lt 1-{\m\over r}\rt^{-1} dr^2+  H^{3\over2} r^2 d\Omega^2_{2},
\nonumber\\
e^{2\f}&=& H^{1\over 2},\q H=1+{\m \sinh ^{2} \a \over r},
\eea
where for simplicity we have set the  constant mode of the 4d 
dilaton $\f_{0}=0$.
The integration constants $\mu$ and $\a$ are related to the mass and 
charge of the solution by
\be\lb{h3}
M= {1\over 2} \mu \lt 1+{3\over 2}\sinh ^{2} \a\rt, 
\q Q={1\over 2} \sinh 2 \a.
\ee
To be more precise, $Q$ is the (common) charge  of the three $U(1)$ 
fields in the action (\ref{e1}). $Q$ is related to the charge $\hat Q$ 
of the single $U(1)$ field appearing in the action (\ref{h1}) by $Q= 
(2/\sqrt 3)\hat Q $.
Using  the area law we find that the Bekenstein-Hawking entropy of the hole 
is given by
\be\lb{h4}
S={{\cal A}  \over 4}= \sqrt \mu\lt \mu+ \mu\sinh ^{2} \a\rt^{3/2},
\ee
where ${\cal A}$ is the area of the event horizon.

The extremal black hole is obtained in the limit $\mu\to 0,\, \a \to 
\infty$,  keeping $\mu \sinh ^{2} \a=Q$. In this limit the solution is 
given by Eq. (\ref {h2}) with $\mu=0$ and $H=(1+ Q/r)$ whereas the mass 
is $M_{ex}= {3\over 4}Q$. 
Let us now consider small excitations near the extremal solution, the 
entropy of these configurations is  given by
\be\lb {h4a}
  S_{(4)}= \pi Q^{3/2} \sqrt 
 \mu 
+\ord{\m^{3/2}}.
\ee

We know that, when expressed in terms of the string metric, the near-horizon, 
near-extremal, magnetically charged,   black hole solution (\ref{h2})
factorizes as ${\cal M}^{2}\times S^{2}$, where  ${\cal M}^{2}$ is the  
solution of the 2d dilaton gravity model (\ref{f8}) with $N=3,\, m=2, 
a=1/\sqrt 3$ ,
\be\lb{h5}
A={1\over 2}\dxd \eF\lt R+2\la \rt.
\ee  
(We have rescaled $\la \to 2\la$ in order to mach the conventions of 
Ref. \cite{CM}).

The dilaton gravity model (\ref{h5}) admits solutions that can be 
interpreted as  black holes in 2d AdS space \cite{CM2},
\be\lb{h6}
\ds-(\l^2x^2-b^2)dt^2+(\l^2x^2-b^2)^{-1}dx^2,\qquad \eF=e^{-2\F_{0}} \l x.
\ee
The mass $M_{(2)}$ and the entropy $S_{(2)}$ of the 2d black hole are 
given in terms of the integration constants $b, \F_{0}$, by
\be\lb{h7}
M_{(2)}={1\over 2} e^{-2\F_{0}} b^2\l, \q S_{(2)}=4\pi \sqrt{ e^{-2\F_{0}}
 M_{(2)}\over 2\l}.
\ee
If the 2d model (\ref{h5})  has to describe the near-extremal, 
near-horizon regime 
of the 4d black hole solutions (\ref{h2}) then the 2d expression for 
the entropy (\ref{h7}) should match the leading order of the 
corresponding 4d quantity in Eq. (\ref{h4a}). 
To show this, we first write the 4d solution (\ref{h2}) in terms of the string 
metric, we  expand the solution near extremality and  near the horizon.
After some manipulations 
we get 

\bea\lb{h7a}
ds^{2}&=&-\lt\la x^{2}  -\l \mu \rt dt^{2}+ 
\lt\la x^{2} -\l \mu \rt^{-1} dx^2+ e^{2\f_{0}}Q^{2} d\Omega^2_{2},
\nonumber\\
e^{-2\f}&=&\sqrt 2 \l x.
\eea 
 where $\l = 1/(2Q)$.
 
As expected the 4d solution factorizes as the product of a 2d spacetime 
and a two-sphere of constant radius. 
Taking into account that the dimensional reduction $4d\to 2d$ implies 
the following relation between the 2d dilaton $\F$  and the 4d one 
$\f$:
\be\lb{h8}
\eF= {1\over 2} Q^{2} \ef,
\ee
and comparing Eq. (\ref{h7a}) with Eq. (\ref{h6}), we get
\be\lb{h8a}
\mu={2M_{(2)}\over \la} e^{2\F_{0}},\q e^{-2\F_{0}}={\sqrt 2\over 
8\la}.
\ee
Using Eqs. (\ref{h8a}) into the expression (\ref{h4a})  for 
the 4d entropy and taking into account only the leading 
term, we obtain a complete 
agreement  with the 2d results, i.e $S_{(4)}=S_{(2)}$.

Until now we have considered only 4d, magnetically charged, solutions, 
i.e the state $S(3,2)$. The  4d electrically charged solution $S(3,1)$
does not factorize as 
direct product of  two 2d spaces. Near extremality it  is described by the 
2d model with $N=3, m=1$, which  is dual to the model (\ref {h5}).
Because the 2d entropy does not change under Weyl rescaling of the 
metric, it follows that the Eqs. (\ref{h7}) and the equality  
 $S_{(4)}=S_{(2)}$ hold also for excitations 
near 4d extremal, electrically charged, solutions.

We have shown that the semiclassical dynamics of small excitations near 
extremality of the 4d black hole can be described by the 2d model 
(\ref{h5}) and that at the leading order,  the 2d and 4d thermodynamical 
entropy is the same.
One can therefore use the results of Ref. 
\cite{CM} as an indirect calculation of the statistical entropy of the 4d 
black hole (\ref{h2}) in the near-extremal regime.

In Ref. \cite{CM} we have found  a mismatch of a $\sqrt 2$ factor between the 
thermodynamical and the statistical entropy of the 2d black hole. 
This implies that also the 
statistical entropy of the 4d $a=1/\sqrt 3$ black hole agrees only up 
to a $\sqrt 2$ factor with the thermodynamical result. A simple 
explanation of this $\sqrt 2$ factor could be found  when the 
2d AdS black hole  arises as compactification of 3d one \cite{CM}.
The same arguments of Ref. \cite{CM} apply also in the case under 
consideration 
because 
the 4d black hole solution (\ref{h2}) arises as 
compactification of the solutions of   5d Einstein-Maxwell gravity 
that behave, near the horizon, as   AdS$_{3}\times S^{2}$.

Let us consider the 5d Einstein-Maxwell action,
\be\lb{h9}
A=\dxc \lg R- {1\over 4}  F^{2}\rg.
\ee
Compactifying the fifth dimension $x_{4}$, using the ansatz
\bea\lb{l1}
ds^{2}_{(5)}&=&e^{-{4\over \sqrt 3} \hat \f}dx_{4}^{2}+e^{{2\over \sqrt 3} 
\hat \f}ds^{2}_{(4)},\nonumber\\
F_{\hat \mu, \hat \nu}&=&\lg F_{\mu \nu},  F_{4 \nu}\rt, F_{4 \nu}=0,
\q  \hat \mu, \hat \nu=0\ldots 4,
\eea
we get the 4d action (\ref{h1}). 
The extremal 5d solutions (\ref{l1}) behave  near the horizon as 
AdS$_{3}\times S^{2}$,
\be\lb{l2}
ds^{2}_{(5)}={r\over Q}\lt -dt^{2}+ dx_{4}^{2}\rt + \lt{Q\over 
r}\rt^{2}dr^{2}
+Q^{2} d\Omega^2_{2}.
\ee
Hence, the explanation of the $\sqrt 2$ factor of Ref. \cite{CM} can be 
immediately translated to the case under consideration.

Moreover, the expression (\ref{l2}) suggests that the discrepancy 
between
thermodynamical and  statistical entropy of the 4d, $a=1/\sqrt3$, black 
hole could also have a geometrical explanation.  
The 3d  part of the metric (\ref{l2}) describes  a spacetime that is 
AdS$_{3}$ with a conical singularity. In fact, if in Eq. (\ref{l2})  
$ x_{4}= Q\varphi$, with $0\le\varphi\le 2\pi$, changing coordinates 
$r\to r^{2}/4, \varphi \to 2 \varphi$, the 3d part of the metric   
(\ref{l2}) becomes 
\be\lb{l3}
ds^{2}_{(3)}= - {r^{2}\over 4 Q^{2}} dt^{2}+{4 Q^{2}\over r^{2}} 
dr^{2}+r^{2}d \varphi^{2},
\ee
but with $0\le\varphi\le \pi$.

\section {Conclusions}

The dimensional reduction of 4d  heterotic string theory presented in 
this paper has shown once again that 2d dilaton gravity models can be 
used as 
a simplified description that retain the relevant information about 
the 4d physics. The class of 2d dilaton gravity models 
we have derived  gives   a general description 
 of  excitations near the extremal 4d heterotic  black hole.
The geometrical structures, thermodynamical features and the duality 
symmetries  of the 4d  theory  become much simpler when translated in 
the 2d context. 

Particularly interesting are those 2d models that admit $AdS_{2}$ as 
solution. In this case one can use the AdS/CFT duality to compute the 
statistical entropy of the near-extreme 4d black hole. We have performed this 
calculation for the $a=1/\sqrt 3$ black hole but in principle the same 
should be possible for the $a=0$ black hole. 
The near-horizon geometry factorizes also  for $a=0$ as AdS$_{2} 
\times S^{2}$ and the excitations near extremality are now described 
by the model (\ref{f9}). Differently from the $a=1/\sqrt 3$
case, where we have a linear varying $\exp(-2\F)$, for  $a=0$  the dilaton is 
constant.  
A constant dilaton makes a black hole interpretation of the solutions very 
difficult,
at least from the 2d point of view. One cannot use the 
arguments of Ref. \cite{CM} to compute the statistical entropy of the 
black hole.

This difficulties  of the $a=0$ case (actually the most 
interesting case from the string point of view) are connected 
with a peculiarity,  mentioned in the introduction, of the $a=0$ case,
namely the existence of a mass gap separating the extremal configuration 
from the continuous part of the spectrum. This implies that the 
finite-energy excitations near extremality are suppressed 
\cite{MMS}. 
Probably, this behavior is related with  other puzzling features 
of the AdS$_{2}$/CFT$_{1}$ correspondence \cite{MMS,st1, CM}.

 .


\end{document}